\documentclass[sigconf,screen]{acmart}
\AtBeginDocument{%
  }

\setcopyright{cc}
\setcctype[4.0]{by}

\copyrightyear{2025}
\acmYear{2025}
\acmConference[PROMISE '25]{Proceedings of the 21st International Conference on Predictive Models and Data Analytics in Software Engineering}{June 2025}{Trondheim, Norway}

\acmBooktitle{Proceedings of the 21st International Conference on Predictive Models and Data Analytics in Software Engineering (PROMISE '25), June 2025, Trondheim, Norway}

\acmDOI{}
\acmISBN{}

\makeatletter
\def\@copyrightspace{\relax}
\makeatother

\usepackage{comment}
\usepackage{array}
\usepackage{tikz}
\usepackage{float}
\usepackage{tikz}
\usepackage{hyperref}

\newcommand*\circled[1]{\tikz[baseline=(char.base)]{
            \node[shape=circle,draw,inner sep=.7pt] (char) {#1};}}

\begin{document}

\title{Towards Build Optimization Using Digital Twins}

\author{Henri A\"idasso}
\email{henri.aidasso.1@ens.etsmtl.ca}
\orcid{0009-0004-1625-0159}
\affiliation{%
  \institution{École de technologie supérieure}
  \streetaddress{1100 Notre-Dame Street West}
  \city{Montréal}
  \state{Québec}
  \country{Canada}
}

\author{Francis Bordeleau}
\email{francis.bordeleau@etsmtl.ca}
\orcid{0000-0001-7727-3902}
\affiliation{%
  \institution{École de technologie supérieure}
  \streetaddress{1100 Notre-Dame Street West}
  \city{Montréal}
  \state{Québec}
  \country{Canada}
}

\author{Ali Tizghadam}
\email{ali.tizghadam@telus.com}
\orcid{0000-0002-0898-3094}
\affiliation{%
  \institution{TELUS}
  \city{Toronto}
  \state{Ontario}
  \country{Canada}}

\begin{abstract}
 Despite the indisputable benefits of Continuous Integration (CI) pipelines (or builds), CI still presents significant challenges regarding long durations, failures, and flakiness. Prior studies addressed CI challenges in isolation, yet these issues are interrelated and require a holistic approach for effective optimization. To bridge this gap, this paper proposes a novel idea of developing Digital Twins (DTs) of build processes to enable global and continuous improvement. To support such an idea, we introduce the CI Build process Digital Twin (CBDT) framework as a minimum viable product. This framework offers digital shadowing functionalities, including real-time build data acquisition and continuous monitoring of build process performance metrics. Furthermore, we discuss guidelines and challenges in the practical implementation of CBDTs, including (1) modeling different aspects of the build process using Machine Learning, 
(2) exploring what-if scenarios based on historical patterns, and (3) implementing prescriptive services such as automated failure and performance repair to continuously improve build processes.

\end{abstract}




\maketitle

\section{Introduction}
The use and evolution of continuous integration and continuous deployment (CI/CD) pipelines -- also known as builds -- come with several challenges that prior research addressed isolately. CI challenges are mainly related to long build duration, frequent build failures, and flaky (non-deterministic) builds \cite{aidasso_build_2025, hilton_trade-offs_2017}. These issues incur significant waste to organizations, including high CI costs, reduced developer productivity, and delayed software releases \cite{hilton_trade-offs_2017, lampel_when_2021, olewicki_towards_2022}. To address CI challenges, prior studies have proposed techniques for improving specific build performance metrics in isolation. For instance, \citet{gallaba_accelerating_2022} investigated strategies to accelerate build duration, \citet{lou_history-driven_2019} developed an automated repair technique to reduce build failures, while \citet{olewicki_towards_2022} focused on detecting flaky builds to minimize costs associated to build flakiness. 

Despite the proven benefits of the CI-improving solutions, practitioners remain reluctant to adopt them due to uncertainties associated with build process changes \cite{senapathi_devops_2018}. In fact, build performance metrics (e.g. duration, failure, flakiness) are interrelated in such a way that process changes made to improve one may inadvertently degrade another.
As evidence, \citet{ghaleb_studying_2023} found that while enabling cache may speed up build duration, it can also lead to failures due to dependencies becoming obsolete.
Hence, there is a strong need for practitioners to \textbf{(1) monitor} the initially sought performance improvement when adopting a solution, 
\textbf{(2) analyze} the impact of process changes on other performance metrics, and \textbf{(3) optimize} multiple performance metrics in a single framework. Such challenges align with Lehman's law stipulating that software systems are "\textit{...multi-level, multi-loop, multi-agent feedback systems and must be treated as such to achieve significant improvement...}" \cite{lehman_understanding_1979}. 
Yet, to our knowledge, no prior study has approached the continuous improvement of CI build processes from a holistic standpoint, which makes the need to fill this gap long overdue.

\begin{figure}[ht]
    \centering
    \includegraphics[width=.35\textwidth]{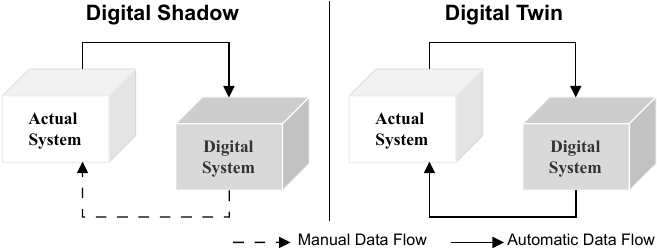}
\caption{Concepts of Digital Shadow and Digital Twin.}
\label{fig:DSvsDT}
    \Description{Concepts of Digital Shadow and Digital Twin.}
\end{figure}

Digital Twins (DTs) have emerged as a key technology that enables real-time monitoring and optimization of complex systems, referred to as the Actual Twin (AT). \citet{glaessgen_digital_2012} defined a DT as a multiscale and dynamic representation of a complex system whose functions are to mirror the life of its corresponding AT. As such, the synchronization between the DT and its AT is fundamental and is achieved through the automated integration of real-time data from the AT. As illustrated in Fig.~\ref{fig:DSvsDT}, this level of implementation forms the Digital Shadow (DS), which supports continuous monitoring and enables manual improvement feedback to the AT. In addition, the DT includes mathematical models that provide intelligent capacities used to automate the feedback loop for improvement of the AT. As a result of their effectiveness, DTs have been used to optimize a broad range of cyber-physical systems such as aircraft, autonomous vehicles, wind farms, and manufacturing processes. 
Furthermore, many DT frameworks such as Eclipse Ditto have been developed and open-sourced to facilitate the implementation of DTs, but mainly in IoT \cite{gil_survey_2024}.

This paper presents the following contributions: (1) a novel idea of developing CI Build process Digital Twins (CBDTs) to continuously monitor and improve continuous integration build processes across multiple performance metrics, (2) a context-agnostic framework with digital shadowing capabilities to demonstrate the feasibility of this idea and facilitate CBDTs' development, and (3) guidelines and challenges for developing CBDTs in practice. A replication package including the framework's source code is made available. \href{https://doi.org/10.6084/m9.figshare.27641388}{\includegraphics[width=4.8cm]{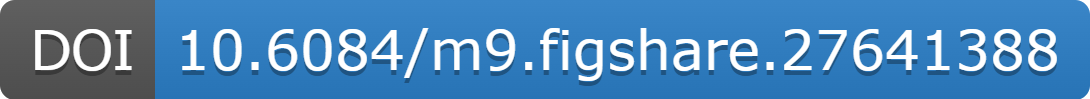}}

\begin{figure}[t!]
    \advance\leftskip-.36cm
    \includegraphics[width=.5\textwidth]{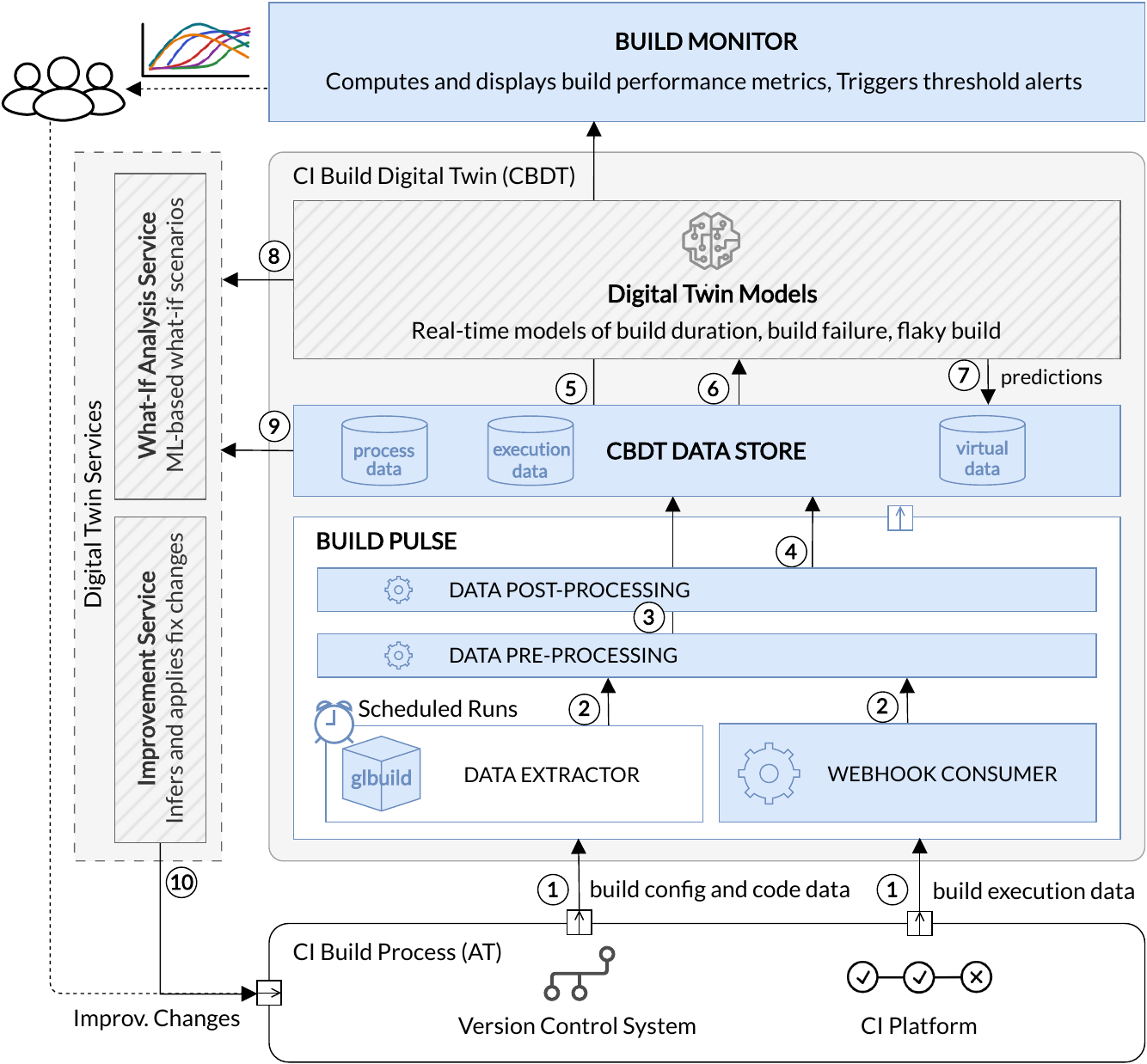}
\caption{Architecture of the CI Build process Digital Twin}
\label{fig:cbdt}
    \Description{A diagram showing the architecture of the CI Build Process Digital Twin, including its components and interactions.}
\end{figure}

\section{CI Build Digital Twin -- CBDT}

Fig.~\ref{fig:cbdt} illustrates the architecture of the proposed CBDT for continuous monitoring and global improvement of the CI Build Process (AT). In organizations, this AT mainly encompasses the Version Control System (VCS) and the associated CI Platform.  The objectives and conceptual design of the CBDT are presented below.

\subsection{Optimization Objectives}
\label{sec:objectives}

A DT is developed with clearly defined optimization objectives, i.e. key performance metrics that will guide optimization efforts. Based on a recent systematic literature review \cite{aidasso_build_2025}, we identify four primary performance metrics that should be all minimized for global build optimization:
    \textbf{Executions Frequency} -- A higher number of build executions indicates more frequent software delivery but also higher build costs.
    \textbf{Mean Duration} -- A lower value indicates faster feedback, which is key for productivity. An optimal baseline of build duration is $\approx$ 10 min \cite{hilton_trade-offs_2017}.
    \textbf{Failure Ratio} -- The ratio of failed jobs to completed jobs. A lower FR is better as failures take up diagnosis time, slowing software delivery. 
    \textbf{Flaky Failure Ratio} -- Percentage of failed jobs that changed to success after retries. A higher ratio points to environmental instabilities causing waste.
    More specific proxies of these metrics (e.g. execution cost) are to be defined to correspond to contextual development objectives.

\subsection{Interactions and Collaboration}

This section presents our conception of how the different components of the CBDT interact with the build process (AT) to enable its continuous monitoring and global improvement. 

Fig.~\ref{fig:cbdt} highlights arrows annotated from \circled{1} to \circled{10}, indicating the main sequence of interactions between the CBDT components. First, data collection tools (in \textsc{Build Pulse}) continuously observe the AT via webhook events, and \circled{1} collect raw build execution and code-related data using APIs each time the AT state changes. The collected data are \circled{2} pre-processed before being \circled{3} stored in the \textsc{CBDT Data Store}.  
After data integration, additional data fields required for ML models and visualization are \circled{4} calculated at the post-processing stage and stored. Every time new build data are integrated, the newly available data are \circled{5} queried by the monitoring service \textsc{Build Monitor} to refresh the visualizations, and \circled{6} by the DT Models for continual learning \cite{wang_comprehensive_2024} of the ML models. These ML models make real-time predictions, which are also \circled{7} stored for monitoring and analytics purposes discussed in Section~\ref{sec:build_cbdt}. Furthermore, \circled{8} replicas of the models and \circled{9} the CBDT data are used to provide the additional DT services, including data-centric what-if analysis \cite{grafberger_towards_2022} and improvement services such as ML-based build repair techniques that infer patches from historical failure repair patterns \cite{lou_history-driven_2019}. It is worth noting that data flows \circled{6} and \circled{9} follow a publish-subscribe messaging pattern, where \textsc{Build Pulse} acts as the publisher, notifying data subscriber components via the \textsc{CBDT Message Broker} each time new data is integrated. For simplicity, this message broker is not illustrated. Finally, \circled{10} the improvement changes (e.g., code patches) are automatically applied to the AT through its APIs for continuous improvement.

\section{CBDT Framework}

To validate the feasibility of our concept, we developed the CBDT framework as a minimum viable product (MVP) composed of the \textcolor{cyan}{\textbf{blue}} components in Fig.~\ref{fig:cbdt}. The basic features of these components are context-agnostic and can be easily reused for different specific CBDT development objectives. The remaining components in \textbf{gray}, require contextual implementations discussed in Sections~\ref{sec:build_cbdt} and \ref{sec:optimization}.

At a minimum, researchers and practitioners can use the CBDT framework to obtain build data for their predictive models and analytics use cases. Unlike historical build data collection tools like TravisTorrent \cite{beller_travistorrent_2017}, this framework incorporates real-time data acquisition and distribution capabilities, facilitating the development of real-time prediction models. In its intended form, the CBDT framework provides digital shadow functionalities (i.e. real-time data synchronization and continuous monitoring) and forms the backbone for implementing and using CBDT models and services in diverse software project contexts.

\subsection{Implementation Setup} To implement the CBDT framework, we focused on GitLab, a unified DevOps platform that incorporates both VCS and CI Platform and is used in many organizations worldwide. We verified the CBDT framework's functionalities by digitally shadowing the build process of the 10 most popular GitLab projects over the second half of 2024. As a result, we accumulated 25GB of data from 1.7 million build jobs. This dataset, along with the source code of the CBDT framework --- including \texttt{Docker} images and \texttt{docker-compose.yaml} files to run the framework --- are made publicly available in the replication package to foster reuse, and development of CBDTs.

\subsection{Framework Components} The CBDT framework is composed of five main microservices.

 \subsubsection{Build Pulse}
 \label{sec:build_pulse}
 
It is the core Python application responsible for synchronizing data with the AT and distributing this data to the other CBDT components. 
 To achieve data synchronization, it includes modules for historical and real-time build data acquisition.

  Historical data integration is necessary to get a baseline of past performance and behaviors. For this purpose, we developed a lightweight Python library {\textit{glbuild}} that collects historical data from a set of GitLab projects. It retrieves and stores data from the complete build history or until it reaches a configurable limit of historical build jobs to save storage space\footnote{There is a wealth of build data in very active and old projects, so this option helps to avoid time-consuming collection of oldest jobs and help save storage space.}. In addition, \textit{glbuild} accepts an object parameter implementing the abstract class \texttt{AbstractStorage}, which handles data pre-processing and storage during collection. Pre-processing steps involve parsing raw data to defined data models, data cleaning and formatting (e.g., uniforming timestamps to UTC timezone). Post-processing steps are run to compute additional data fields for ML trainings. Currently, only the \texttt{flaky} field value is computed based on non-deterministic job reruns \cite{olewicki_towards_2022}.

Real-time data integration ensures that the CBDT remains synchronized with the AT. For this purpose, we implemented a webhook consumer endpoint to listen to the job webhook events\footnote{\url{https://docs.gitlab.com/user/project/integrations/webhook_events}} triggered every time a build job status changes. The consumer endpoint is secured using a token-based middleware to ensure that data is received only from the GitLab webhooks. Once the consumer receives a job event, API calls are made using the job ID to collect all the required build data. The obtained data then undergo the same pre-processing steps described for historical data before being stored, and the same post-processing steps afterward. Configuring job webhook events in GitLab requires administrator access to the projects. To circumvent this requirement when working on open-source projects where such access is not available, we implemented scheduled integrations of historical data as illustrated in Fig.~\ref{fig:cbdt}. These scheduled runs are activated by configuring the \texttt{DATA\_REFRESH\_INTERVAL} environment variable in \textsc{Build Pulse}, which specifies the time interval in seconds at which \textit{glbuild} reruns to fetch the newly available build job data since its last run.

Each time \textsc{Build Pulse} integrates newly available build data, it notifies all the data subscriber components via the \textsc{CBDT Message Broker}. It also exposes REST APIs for searching through the integrated data using criteria on various data fields.

\subsubsection{Build Monitor} This is the data visualization application enabling real-time build process monitoring. For its implementation, we extended Grafana OSS\footnote{\url{https://grafana.com/oss/grafana}} with provisioned data source, dynamic SQL queries, and dashboard setups for real-time visualization of the build performance metrics defined in Section~\ref{sec:objectives}. Via the queries, these performance metrics are calculated and refreshed on the dashboard at a configurable time interval. The monitoring dashboard enables metrics aggregation at different time intervals (i.e., hourly, daily, weekly, monthly, and yearly) and data granularity levels (i.e., single versus multiple projects). \textsc{Build Monitor} also enables threshold-based alerts (e.g., e-mail), allowing practitioners to take prompt action to address issues before they escalate.
    
\subsubsection{Build Data Subscriber} It is the boilerplate Python application designed to subscribe to DT data updates. It includes minimal functions required to listen to data availability signals that \textsc{Build Pulse} emits via the \textsc{CBDT Message Broker}. This boilerplate application serves as a foundation for building contextual data subscriber components (with gray and hashed background in Fig~\ref{fig:cbdt}) that require build data stream to synchronize their states and automate actions.

\subsubsection{CBDT Data Store} It is responsible for storing and providing access to the DT data. We use PostgreSQL, an open-source database system that efficiently stores large volumes of structured and unstructured data and offers high querying performance. While the current CBDT framework version uses PostgreSQL, the framework can easily be customized to interact with other storage solutions --- by implementing a concrete definition of the \texttt{AbstractStorage} class mentioned in Section~\ref{sec:build_pulse}.

\subsubsection{CBDT Message Broker} It is the publish-subscribe software used to notify the data subscribers of the availability of new build data from \textsc{Build Pulse}. For simplicity, this broker is not explicitly shown in the CBDT architecture. We use RabbitMQ\footnote{\url{https://www.rabbitmq.com}}, a very popular message broker that offers persistence and high availability.

\section{Leveraging Machine Learning to Build CBDTs}
\label{sec:build_cbdt}

This section discusses guidelines for creating the DT models essential for \textbf{real-time predictions and anomaly detection}.

Based on the target performance metrics defined in Section~\ref{sec:objectives}, we propose creating ML models of three key build process aspects: failure, duration, and flakiness. Beyond the generally intended use cases of these models in the literature --- i.e. assist developers in managing their schedules and proactively tackling failures \cite{aidasso_build_2025} --- the continuous monitoring of the models' outcomes and actual states can help practitioners better understand the models' decisions while enabling real-time anomaly detection by identifying deviations from predicted (expected) states. Concretely, the idea is to have real-time monitoring of the predicted and actual states so that sudden spikes in actual failures, duration, or flakiness (compared to predictions) could be easily linked to recent process changes. Similarly, improvements resulting from process changes can be measured by a decline in these metrics.

To create the DT models, one can leverage existing techniques for build failure prediction, duration prediction, and flaky build detection reported in \cite{aidasso_build_2025}. Algorithms used in state-of-the-art techniques mainly include tree-based models like XGBoost and Random Forest, which offer easy interpretability, as well as deep learning models like Long Short-Term Memory (LSTM) networks, which effectively capture temporal patterns in data. In addition, modeling build processes involves transforming collected data into features associated with code changes, CI performance, infrastructure telemetry, developers and project-specific characteristics, as reported in \cite{aidasso_build_2025}. 

The challenge of leveraging ML in the context of CBDTs is two-fold: (1) creating several multi-scale models within the same framework, including project-specific models, and cross-project models that leverage data from other projects to improve predictions within a single project; and (2) continuously using the numerous real-time models while maintaining high availability and minimal latency, and updating them online (e.g., \cite{ni_acona_2018}) to avoid concept drift.

\section{Build Optimization with the CBDT}
\label{sec:optimization}

\subsection{What-if Analysis Service}

In the context of CBDTs, what-if analyses refer to exploring how different build process changes (i.e. scenarios) impact each of the build performance metrics. Such analyses will enable practitioners to explore alternative scenarios and make informed decisions by identifying process changes that maximize benefits while minimizing negative impacts on other performance metrics. The what-if analysis service can be implemented using ML models as discussed in \cite{grafberger_towards_2022}. 
The idea is to use replicas of the DT models to evaluate the sensitivity of the build performance metrics to specific changes in the build configuration (e.g., enabling cache or updating the testing framework version). To achieve this, DT models should integrate process-related features, such as numerical representations of build configuration files (e.g., \textit{.gitlab-ci.yml}), which can be obtained as embeddings from language models. Moreover, effective what-if analysis requires learning from large enough historical build data from various projects to reliably estimate how process modifications influence the different build performance metrics.

\subsection{Improvement Services} 
Various services can be implemented to infer improving changes and automate actions for the continuous improvement of the AT. 
For example, when the DT duration model predicts a long build time, build acceleration techniques like caching \cite{gallaba_accelerating_2022} can be applied. If a build failure is predicted, ML-based automated repair techniques such as \cite{lou_history-driven_2019} can be used to infer and apply patch candidates via the AT's update API endpoints (e.g, Repository files API\footnote{\url{https://docs.gitlab.com/ee/api/repository_files.html}} and CI/CD variables API\footnote{\url{https://docs.gitlab.com/ee/api/instance_level_ci_variables.html}}). While efforts have been made toward the diagnosis of flaky builds \cite{lampel_when_2021}, techniques for automating the repair of such builds are yet to be developed to minimize the associated waste. Overall, the envisioned idea is that such recurring and automated improvement actions, executed in concert with the DT models' predictions and informed decision-making about process changes, will ensure continuous and global improvement of build processes.

\section{Conclusion}

This paper introduces a novel idea of developing CI Build Digital Twins (CBDTs) for holistic and continuous optimization of build processes. Essentially based on continuous data mining, the CBDT aims to provide various services including real-time monitoring, anomaly detection, what-if analysis, and automated improvement of multiple build process performance metrics including failures, duration, and flakiness. To support such an idea, we developed the CBDT framework which in its current state enables digital shadowing. This MVP framework is part of ongoing research work toward developing CBDTs with industrial partners who see great value in this DT-based approach for global build optimization.

\bibliographystyle{ACM-Reference-Format}
\bibliography{references}

\end{document}